\documentclass[12pt]{article}
\usepackage{amsmath}
\usepackage{epsf}
\DeclareSymbolFont{boldletters}{OML}{cmm} {b}{it}
\DeclareSymbolFontAlphabet{\mathbit}{boldletters}
\DeclareMathSymbol{\alpha}{\mathalpha}{letters}{"0B}
\DeclareMathSymbol{\beta}{\mathalpha}{letters}{"0C}
\DeclareMathSymbol{\gamma}{\mathalpha}{letters}{"0D}
\DeclareMathSymbol{\delta}{\mathalpha}{letters}{"0E}
\DeclareMathSymbol{\epsilon}{\mathalpha}{letters}{"0F}
\DeclareMathSymbol{\zeta}{\mathalpha}{letters}{"10}
\DeclareMathSymbol{\eta}{\mathalpha}{letters}{"11}
\DeclareMathSymbol{\theta}{\mathalpha}{letters}{"12}
\DeclareMathSymbol{\iota}{\mathalpha}{letters}{"13}
\DeclareMathSymbol{\kappa}{\mathalpha}{letters}{"14}
\DeclareMathSymbol{\lambda}{\mathalpha}{letters}{"15}
\DeclareMathSymbol{\mu}{\mathalpha}{letters}{"16}
\DeclareMathSymbol{\nu}{\mathalpha}{letters}{"17}
\DeclareMathSymbol{\xi}{\mathalpha}{letters}{"18}
\DeclareMathSymbol{\pi}{\mathalpha}{letters}{"19}
\DeclareMathSymbol{\rho}{\mathalpha}{letters}{"1A}
\DeclareMathSymbol{\sigma}{\mathalpha}{letters}{"1B}
\DeclareMathSymbol{\tau}{\mathalpha}{letters}{"1C}
\DeclareMathSymbol{\upsilon}{\mathalpha}{letters}{"1D}
\DeclareMathSymbol{\phi}{\mathalpha}{letters}{"1E}
\DeclareMathSymbol{\chi}{\mathalpha}{letters}{"1F}
\DeclareMathSymbol{\psi}{\mathalpha}{letters}{"20}
\DeclareMathSymbol{\omega}{\mathalpha}{letters}{"21}
\DeclareMathSymbol{\varepsilon}{\mathalpha}{letters}{"22}
\DeclareMathSymbol{\vartheta}{\mathalpha}{letters}{"23}
\DeclareMathSymbol{\varpi}{\mathalpha}{letters}{"24}
\DeclareMathSymbol{\varrho}{\mathalpha}{letters}{"25}
\DeclareMathSymbol{\varsigma}{\mathalpha}{letters}{"26}
\DeclareMathSymbol{\varphi}{\mathalpha}{letters}{"27}
\DeclareMathSymbol{\Gamma}{\mathalpha}{letters}{"00}
\DeclareMathSymbol{\Delta}{\mathalpha}{letters}{"01}
\DeclareMathSymbol{\Theta}{\mathalpha}{letters}{"02}
\DeclareMathSymbol{\Lambda}{\mathalpha}{letters}{"03}
\DeclareMathSymbol{\Xi}{\mathalpha}{letters}{"04}
\DeclareMathSymbol{\Pi}{\mathalpha}{letters}{"05}
\DeclareMathSymbol{\Sigma}{\mathalpha}{letters}{"06}
\DeclareMathSymbol{\Upsilon}{\mathalpha}{letters}{"07}
\DeclareMathSymbol{\Phi}{\mathalpha}{letters}{"08}
\DeclareMathSymbol{\Psi}{\mathalpha}{letters}{"09}
\DeclareMathSymbol{\Omega}{\mathalpha}{letters}{"0A}
\newcommand{\mbit}[1]{{\mathbit#1}}

\newcommand{\dsl}[2]{{#1}{\mbox{\hspace{-8pt}\hspace{#2}$\not$}
\hspace{8pt}\hspace{-#2}}} 

\newcommand{\dslp}{\dsl{p}{1.5pt}}

\newcommand{\dsll}{\dsl{l}{.5pt}}
\newcommand{\dslpar}{\dsl{\partial}{0pt}}

\begin{document}

\title{
\begin{flushright}
\small{
Ehime-th-2 \\
KYUSHU-HET-67 }
\end{flushright}
Auxiliary Field Method in $4$- and $3$- dimensional Nambu--Jona-Lasinio Models}

\author{Taro~Kashiwa\thanks{kashiwa@phys.sci.ehime-u.ac.jp}\\  
Department of Physics, Ehime University, Matsuyama 790-8577, Japan \\
\\
Tomohiko~Sakaguchi\thanks{tomohiko@higgs.phys.kyushu-u.ac.jp}
\\
Department of Physics, Kyushu University, Fukuoka 812-8581, Japan\\\\}

\date{\today}

\maketitle

\abstract{In order to check the validity of auxiliary field method in the Nambu--Jona-Lasinio model, the one-loop (=quantum) effects of auxiliary fields to the gap equation are considered   with $N$-component fermion models in $4$ and $3$ dimensions. 
$N$ is not assumed so large but regarded as a loop expansion parameter. 
To overcome infrared divergences caused by the Nambu-Goldstone bosons, an intrinsic fermion mass is assumed. It is shown that the loop expansion can be justified by this intrinsic mass 
whose lower limit is also given. It is found that due to quantum effects, chiral symmetry breaking ($\chi$SB) is restored in $D=4$ and $D=3$ when the four-Fermi coupling is large. However, $\chi$SB is enhanced in a small coupling region in $D=3$.}

\maketitle\thispagestyle{empty}
\newpage

\section{Introduction}
The Nambu--Jona-Lasinio (NJL) model\cite{rf:NJL} is, needless to say, one of the most famous field theoretical model exhibiting 
dynamical chiral symmetry breaking ($\chi$SB) phenomena. 
Owing to its simplicity, there have been many studies 
using the NJL models: some of recent trends are those which deal 
with external disturbances such as background gauge fields
\cite{rf:KL,rf:GMS,rf:IKT1,rf:IKT2} or curved spacetime\cite{rf:IMO} 
in order to explore the detailed phase structure. 
Originally, calculations for dynamical $\chi$SB had been done 
in a self-consistent manner to yield the gap equation\cite{rf:NJL} 
but later it was revealed that results are much easily obtained 
in path integral with the use of the auxiliary\cite{rf:GNKK} 
or the Hubbard-Stratonovich\cite{rf:FRAD} field. 
The recipe which we shall call the auxiliary field method\cite{rf:KOS} 
becomes exact when $N$, the number of degrees of freedom of the original dynamical fields, goes to infinity. However, the  analysis in lower dimensional bosonic models 
shows that even if $N=1$, we can improve results toward the true value by taking higher orders in the loop expansion\cite{rf:kashiwa}. Therefore it is desirable to incorporate quantum(=loop) effects of 
auxiliary fields to the gap equation when $N$ remains finite.

However, there is an obstacle to perform the loop expansion 
in terms of auxiliary fields in the NJL model: 
due to massless Nambu-Goldstone bosons, members of auxiliary fields, 
infrared divergences are inevitable in higher loop calculations. 
Kleinert and Bossche pay an attention to this infrared regime 
to conclude that there is no pion in the NJL model\cite{rf:kleinert}, 
that is, there is no room for the auxiliary fields. 
Their main ingredient is, however, a chiral nonlinear model\cite{rf:KB}, 
an effective theory, so that there need more rigorous 
and careful investigations. 
Indeed some oppositions to this conclusion have been raised\cite{rf:BLO}.

We follow the standard prescription for the effective action formalism 
by introducing sources coupled to bilinear terms of fermions. 
In order to control the infrared singularity, 
we assume an intrinsic mass of fermions, so called a current quark mass. 
We do not care about the renormalizability so that an ultraviolet cutoff 
is introduced to define a model. 
We will make no approximation other than the loop expansion. 

There have been attempts to consider $\mathcal{O}(1/N)$ terms 
in four-Fermion models: 
some calculate the effective potential for the model with a discrete 
chiral symmetry in order to clarify the renormalizability 
in $4>D>2$\cite{rf:HKWY} , others check the Nambu-Goldstone theorem, 
Gell-Mann--Oakes--Renner or Goldberger-Treiman relations 
in the $4$-dimensional NJL model
\footnote{It is, however, almost trivial to check these relations 
under $\mathcal{O}(1/N)$ expansion; since it is well known that those are 
the consequences of the Ward-Takahashi relations(WTs) which are 
persistent with the loop(=$\mathcal{O}(1/N)$) expansion.} 
to show that $\mathcal{O}(1/N)$ effects weaken quark condensation 
$\langle \overline{q}q \rangle $ obtained by the tree order\cite{rf:NBC}. However, there seems no attempt to study the higher loop contribution, paying an attention to the infrared regime, 
to the vacuum condition and to the gap equation in terms of 
auxiliary field method. In \S2, we present a general formalism in path integral to obtain an effective potential in the NJL model. 
We work with the $N$-component fermion model in $D=4$ and 
$D=3$\cite{rf:ABKW} but as stated above $N$ is merely 
a loop expansion parameter to be kept finite. 
The next section \S3 deals with the vacuum condition and then 
the gap equation up to the one-loop order of the auxiliary fields. 
It is concluded that quantum(=loop) effects restore $\chi$SB in $D=4$ while the situation in $D=3$ is slightly different; $\chi$SB is restored in a strong coupling regime but enhanced in a weak coupling regime. 
We also find the lower limit of a current quark mass 
to ensure the loop expansion. 
The final section is devoted to discussions.

\section{Model and Basic Formalism}

In this section, we develop a general formalism 
in order to clarify our goal: the NJL Model with an intrinsic mass 
$\varepsilon$ in 3 as well as 4 dimensions is given as
\begin{eqnarray} 
{\cal L}& \! \! = \! \! & -\overline\psi(x)
\left( \dslpar  + \varepsilon  \right)  \psi(x)
+ {\cal L}_{\rm int}    \\ 
{\cal L}_{\rm int} & \! \! \equiv \! \! & \frac{\lambda}{2N} 
\left\{\begin{array}{ll}
\left[\bigl( \overline\psi(x)\psi(x) \bigr)^{2}
+\bigl(\overline\psi (x) i\gamma_{5}\psi(x) \bigr)^{2}\right]\ ,& D=4\ ,\\
\noalign{\vspace{1ex}}
\left[\bigl(\overline\psi(x)\psi(x)\bigr)^{2}
+\bigl(\overline\psi(x) i\gamma_{4}\psi(x)\bigr)^{2}
+\bigl(\overline\psi(x) i\gamma_{5}\psi(x)\bigr)^{2}\right]\ , & D=3\ ,
\end{array}\right. \nonumber  \\ 
& \! \! \equiv   \! \! & \frac{\lambda}{2N}\bigl(\overline\psi (x) 
\mbit{\Gamma} \psi(x) \bigr)^{2}  \ ;   \label{GNJL}
\end{eqnarray}
with
\begin{eqnarray}
   \mbit{\Gamma } = \Gamma_a \equiv (1, i \mbit{\Gamma }_5)  \ ;  
\   \mbit{\Gamma }_5 \equiv \left\{
\begin{array}{lr}  \gamma_{5}   \ , &  D=4  \\ 
\noalign{\vspace{1ex}}
(\gamma_{4}, \gamma_{5} )   \ , &  D=3 
\end{array}\right.   \ ,
\end{eqnarray}
where $\dslpar \equiv \gamma_{\mu}\partial_{\mu}$ 
and $N$-component fermion fields have been introduced 
and $\gamma_\mu$'s are $4 \times 4$ matrices, even in 3 dimension
\footnote{In 3 dimension, we need an additional ($N$-component) fermion 
to form a four-component spinor\cite{rf:ABKW, rf:IKT1}
$$
\psi=\left(\begin{array}{c}\psi_{1}\\\psi_{2}\end{array}\right)\ ,\quad
\overline\psi\equiv\psi^{\dag}\gamma_{3}
\equiv\left(\begin{array}{cc}\overline\psi_{1}
&-\overline\psi_{2}\end{array}\right)
\equiv\left(\begin{array}{cc}\psi_{1}^{\dag}\sigma_{3}
&-\psi_{2}^{\dag}\sigma_{3}
\end{array}\right)\ ,
$$
to be able to realize a chiral symmetry, (when $\varepsilon =0$),  
$$
\psi(x)\longrightarrow{\mathrm e}^{i\alpha\gamma_{4}}\psi(x) \ ,  \quad
\psi(x) \longrightarrow{\mathrm e}^{i\beta\gamma_{5}}\psi(x) \ ,
$$
which is, therefore, a global $U(2)$ symmetry, finally broken down 
to $U(1)\times U(1)$ by a mass term.}, satisfying 
$$
\left\{\gamma_\mu , \gamma_\nu \right\} = 2 \delta_{\mu \nu}  \ ; 
\quad \mu, \nu = 1,2,3,4,5 \ .
$$
Explicitly,
\begin{equation}
\gamma_{\mu}=
\left(\! \!  \begin{array}{cc}\sigma_{\mu} \! \! & \! \! 0 \\
0 \! \!  & \! \! -\sigma_{\mu}\end{array} \! \! \right) \ ; 
\ \mu = 1\sim 3   \ ,  \  \ 
\gamma_{4}=\left(\! \!  \begin{array}{cc}0 \! \! & \! \! {\bf1} \\
{\bf1} \! \! & \! \! 0 \end{array} \! \! \right) \ ,  \  \ 
\gamma_{5}=\gamma_{1}\gamma_{2}\gamma_{3}\gamma_{4}
=\left(\! \! \begin{array}{cc}
0 \! \! & \! \! i{\bf1} \\
-i{\bf1} \! \! & \! \!  0 \end{array} \! \! \right) \ .
\label{gamma3d}
\end{equation}
The intrinsic mass $\varepsilon $, so called a current quark mass, 
has been assumed to prevent an infrared divergence 
in pion loop integrals. (See the followings.)

The quantity we should consider is 
\begin{eqnarray}
Z[\mbit{J}]  & \!\! \equiv \!\! & {\rm Tr} \ {\rm T} 
\left(  \exp \left[ - \int_{0}^{T} dt H(t) \right]  \right)  \ ;  \\
H(t) & \!\!  =  \!\! & \int d^{D-1}x \left[ \overline{\psi}(x)
\left( \mbit{\gamma} \cdot \mbit{\nabla } + \varepsilon  
- \mbit{J}(x) \cdot \mbit{\Gamma} \right)  \psi(x) 
-  {\cal L}_{\rm int} - \frac{N}{2 \lambda }\mbit{J}^2(x)  \right]  \ ,  
\label{Hmailtonian} \\
& \!\!   \!\! & \mbit{J }(x)  = {J }_a(x) \equiv     \left\{
\begin{array}{ll} \left( J(x) , J_{5}(x) \right)  \ , &  D=4 \ , \\
\noalign{\vspace{1ex}}
\left( J(x), J_{4}(x), J_{5}(x) \right)   \ , &  D=3 \ ,
\end{array}\right.  \ ; 
\end{eqnarray}
where $\mbit{J}(x)$'s are c-number sources and ${\rm T}$ designates 
the (imaginary) time-ordered product. 
(We have introduced a $\mbit{J}^2$ term just for a later notational 
simplicity\cite{rf:HKWY}.) From this we can extract an energy of 
the ground state by putting $\mbit{J} \mapsto 0$ as well as 
$T \mapsto \infty$. The path integral representation reads 
\begin{eqnarray}
Z[\mbit{J}] = \int d[\psi] d[\overline\psi] \exp 
\left( \int d^D x  \left[ -\overline\psi  
\left( \dslpar + \varepsilon -  \mbit{J}\cdot \mbit{\Gamma} \right)   \psi  
+ \frac{\lambda }{2N}\left( \overline\psi \mbit{\Gamma } \psi \right)^2 
+  \frac{N}{2 \lambda }\mbit{J}^2 \right]   \right) \ .\label{Zoriginal}
\end{eqnarray}
Introducing auxiliary fields in terms of the Gaussian integration,
\begin{eqnarray}
1 & =& \int d[\mbit{\Sigma}] \exp 
\left[ - \frac{N}{2 \lambda } \int d^D x 
\left( \mbit{\Sigma}(x) +  \frac{\lambda}{N}\overline\psi(x) 
\mbit{\Gamma } \psi(x) \right)^2  \right] \ ,  \\  
& &  \mbit{\Sigma }(x) = \Sigma_a(x) \equiv  \left\{
\begin{array}{ll} 
(\sigma(x) , \pi(x)  ) \ , &  D=4 \ , \\
\noalign{\vspace{1ex}}
(\sigma(x) , \pi_1(x) , \pi_2(x) ) \ , &  D=3 \ ,
\end{array}\right.
\end{eqnarray}
to eliminate the four-Fermi interaction, we find
\begin{eqnarray}
Z[\mbit{J}] & = & \int d[\psi] d[\overline\psi] d[\mbit{\Sigma } ] \exp 
\Bigg( \int d^{D}x \Big[- \frac{N}{2 \lambda } \mbit{\Sigma }^2   
\nonumber  \\ 
& & - \overline\psi  \left\{ \dslpar + \varepsilon +  
\left( \mbit{\Sigma} - \mbit{J} \right)  \cdot \mbit{\Gamma }   \right\}  
\psi +  \frac{N}{2 \lambda }\mbit{J}^2  \Big]  \Bigg)   \ . 
\end{eqnarray}
The fermion integrations yield 
\begin{eqnarray}
Z & \hspace{-3mm} = \hspace{-3mm} & \int d[\mbit{\Sigma } ] \exp 
\left[ -N \left\{  \int d^D x \frac{1}{2 \lambda } \left( \mbit{\Sigma }^2 
- \mbit{J}^2 \right)  - {\rm Tr }
\ln \left\{ \dslpar + \varepsilon +  \left( \mbit{\Sigma} - \mbit{J} \right) 
\cdot \mbit{\Gamma }   \right\}  \right\}   \right]  \nonumber  \\ 
& \! \! \! \! \stackrel{\mbit{\Sigma}\rightarrow \mbit{\Sigma}+ \mbit{J}}{ =} 
\! \! \! \! & \int d[\mbit{\Sigma } ] \exp 
\left( -N I[\mbit{\Sigma }, \mbit{J}] \right) \ ,  \label{Zeff} 
\end{eqnarray}
with 
\begin{eqnarray}
I[\mbit{\Sigma }, \mbit{J}]  \equiv    \int d^D x 
\left( \frac{1}{2 \lambda } \mbit{\Sigma }^2 
+ \frac{1}{ \lambda }\mbit{\Sigma }\cdot  \mbit{J} \right)  
- {\rm Tr }\ln \left( \dslpar + \varepsilon +  \mbit{\Sigma}  
\cdot \mbit{\Gamma }   \right)     \  , 
\end{eqnarray}
where Tr designates spinorial as well as functional trace. Write
\begin{eqnarray}
Z[\mbit{J}] \equiv {\mathrm e}^{-N W[\mbit{J}]}  \ , 
\end{eqnarray}
and introduce ``classical" fields,
\begin{eqnarray}
\frac{1}{\lambda }\mbit{\phi} \equiv  \frac{\delta W }{\delta \mbit{J}} 
\stackrel{\mbox{eq.(\ref{Zeff})}}{=} 
\frac{1}{\lambda }<\mbit{\Sigma}> \stackrel{\mbox{eq.(\ref{Zoriginal})}}{ =} 
- \frac{1}{N} < \overline\psi \mbit{\Gamma }  \psi >  
- \frac{1}{\lambda } \mbit{J}   \ , 
\label{quantumField}
\end{eqnarray}
with expectation values being taken under the expression (\ref{Zeff}) 
or under the original one (\ref{Zoriginal}), 
then perform a Legendre transformation with respect to $W[\mbit{J}]$ 
to obtain the effective action,
\begin{eqnarray}
\Gamma[\mbit{\phi}] = W[\mbit{J}] - \frac{1}{\lambda} 
\Big( \mbit{J}\cdot \mbit{\phi} \Big)  \ ; \label{DefEffectiveAction} 
\end{eqnarray}
where a shorthand notation,
\begin{eqnarray}
\Big( A \cdot B \Big) \equiv \int d^Dx \ A(x)B(x)  \ ,
\end{eqnarray}
has been employed. When $\mbit{J}$'s are set to be constants, 
an effective action becomes the effective potential:
\begin{eqnarray}
\Gamma[\mbit{\phi}] \stackrel{\mbit{J} \mapsto \mbox{const.}}
{\Longrightarrow } 
V T {\cal V}(\mbit{\phi})  \ , \label{EffectivePot}
\end{eqnarray}
with $V$ being $(D-1)$-dimensional volume.

We calculate $W[\mbit{J}]$ with the help of the saddle point method
\footnote{The result becomes exact when $N$ goes to infinity, 
which, however, is not the case in this analysis: 
$N$ is a mere expansion parameter that finally be put unity.}: 
first, find the classical solution, $\mbit{\Sigma }_0$, 
\begin{eqnarray}
0 = \left.\frac{\delta I}{\delta \mbit{\Sigma}(x)}\right|_{\mbit{\Sigma }_0} 
= \frac{1}{\lambda }(\mbit{\Sigma}_0 + \mbit{J})(x) - {\rm tr} \mbit{\Gamma } 
S(x,x:\mbit{\Sigma}_0) \ , \label{gapEquation}
\end{eqnarray} 
where $S(x,y:\mbit{\Sigma}_0)$ is a fermion propagator 
under the background fields, 
\begin{eqnarray}
\left( \dslpar + \varepsilon + \mbit{\Sigma}_0(x)  \cdot \mbit{\Gamma }  
\right) S(x,y: \mbit{\Sigma}_0) = \delta (x-y)  \ . \label{Propagator}
\end{eqnarray}
Second, expand $I$ around $\mbit{\Sigma}_0$
\begin{eqnarray}
I= I_0 + \frac{1}{2}\Big( \mbit{I}^{(2)}_0 \cdot 
\left( \mbit{\Sigma } - \mbit{\Sigma }_0\right)^2 \Big)  
+ \frac{1}{3!}\Big( \mbit{I}^{(3)}_0 \cdot \left( \mbit{\Sigma } 
- \mbit{\Sigma }_0\right)^3 \Big) + \cdots  \ ,
\end{eqnarray}
where
\begin{eqnarray}
\mbit{I}^{(n)}_0 & \equiv & \left. \frac{\delta^n I}{\delta \mbit{\Sigma}^n} 
\right|_{\mbit{\Sigma}_0}  \ , \\
 \Big( \mbit{I}^{(n)}_0 \cdot \mbit{\Sigma}^n \Big) & \equiv & 
\int d^D x_1 \cdots d^D x_n \frac{\delta^n I}{\delta \Sigma^{a_1}(x_1) 
\cdots \delta \Sigma^{a_n}(x_n)} 
\Sigma^{a_1} (x_1) \cdots \Sigma^{a_n} (x_n) \ .
\end{eqnarray}
Third, put $(\mbit{\Sigma } - \mbit{\Sigma }_0) \mapsto  
\mbit{\Sigma }/ \sqrt{N} $, then perform the Gaussian integration 
with respect to $\mbit{\Sigma }$ to obtain
\begin{eqnarray}
W[\mbit{J}] = I_0 + \frac{1}{2N} {\rm Tr} \ln \mbit{I}^{(2)}_0 
+ \mathcal{O}(\frac{1}{N^2}) \ ,  \label{ValueofWJ}
\end{eqnarray} 
where
\begin{eqnarray}
I_0 & \! \! =\! \!&  \ \frac{1}{2 \lambda } 
\Big( \mbit{\Sigma}_0 \cdot  \mbit{\Sigma}_0 \Big) 
+  \frac{1}{ \lambda } \Big( \mbit{\Sigma}_0 \cdot \mbit{J}  \Big) 
- {\rm Tr }\ln \left( \dslpar + \varepsilon 
+ \mbit{\Sigma}_0  \cdot \mbit{\Gamma }   \right)  \ ; \label{Izero} \\
\left( \mbit{I}^{(2)}_0 \right)_{ab} & \! \! =\! \!& \left. 
\frac{\delta^2 I}{\delta {\Sigma}^a(x) \delta {\Sigma}^b(y) }
\right|_{\mbit{\Sigma}_0} = \frac{1}{\lambda }\delta (x-y) \delta_{ab} 
+ {\rm tr}{\Gamma }_a S(x,y:\mbit{\Sigma}_0){\Gamma }_b S(y,x:\mbit{\Sigma}_0)
 \ . \label{Itwo}
\end{eqnarray}
Here the trace is taken only for spinor space. 
$\mbit{I}^{(2)}_0$ is a matrix in the $\mbit{\Sigma}$ space 
($2 \times 2$ in $D=4$, $3 \times 3$ in $D=3$). 
$I_0$ is the ``tree" part while ${\rm Tr}\ln \mbit{I}^{(2)}_0$ 
is the ``one-loop" part of the auxiliary fields. 
Using eq.(\ref{quantumField}) we find
\begin{eqnarray}
\mbit{\phi}(x) = \mbit{\Sigma }_0(x) 
+ \frac{\lambda }{2N} \frac{\delta }{\delta \mbit{J}(x)} \left( {\rm Tr} 
\ln \mbit{I}^{(2)}_0 \right)  \equiv  \mbit{\Sigma }_0(x) 
+ \frac{\mbit{\Sigma }_1(x)}{N}  \ . \label{Quantum-ClassicalField}
\end{eqnarray}
Note that difference between $\mbit{\phi}$ and $\mbit{\Sigma }_0$ 
is $\mathcal{O}(1/N)$. Inserting eq.(\ref{Quantum-ClassicalField}) into 
the effective action (\ref{DefEffectiveAction}) with 
the use of eqs.(\ref{ValueofWJ}), (\ref{Izero}), and (\ref{Itwo}), 
we obtain
\begin{eqnarray}
\Gamma [\mbit{\phi}] &\!\! = \!\!& \frac{1}{2 \lambda }
\Big( \mbit{\phi} \cdot \mbit{\phi} \Big)  
- {\rm Tr}\ln \left( \dslpar + \varepsilon + \mbit{\phi} \cdot \mbit{\Gamma}
  \right)    \nonumber  \\ 
& \!\! + \!\! & \frac{1}{2N}{\rm Tr}\ln 
\left( \frac{\bf I}{\lambda }\delta (x-y) 
+ {\rm tr} \mbit{\Gamma}S(x,y :\mbit{\phi} )\mbit{\Gamma}S(y,x :\mbit{\phi})
 \right) + \mathcal{O}(\frac{1}{N^2})\ .
\end{eqnarray}
By putting $\mbit{J}$'s into constants, 
the effective potential (\ref{EffectivePot}) reads
\begin{eqnarray}
{\cal V}(\mbit{\phi}) & \!\! = \!\! & \frac{1}{2 \lambda }\mbit{\phi}^2 
-\frac{1}{VT} {\rm Tr}\ln \left( \dslpar + \varepsilon 
+ \mbit{\phi}\cdot \mbit{\Gamma}  \right)    \nonumber  \\ 
& & + \frac{1}{2NVT}{\rm Tr}\ln \left( \frac{{\bf I}}{\lambda }\delta (x-y) 
+ {\rm tr} \mbit{\Gamma}S(x,y :\mbit{\phi})\mbit{\Gamma}S(y,x :\mbit{\phi}) 
\right) +  \mathcal{O}(\frac{1}{N^2})  \ . \label{effectivePotential}
\end{eqnarray}
In eq.(\ref{effectivePotential}) the first two terms are 
the tree part and the third is the one-loop part 
of the auxiliary fields, whose functional trace should be 
also taken for the $\mbit{\Sigma}$ space. The vacuum is chosen by
\begin{eqnarray}
\left. \frac{\partial {\cal V}}{\partial \mbit{\phi}}\right|_{\mbit{J}=0} = 0
 \ . \label{VacCond}
\end{eqnarray}
Armed with these, we now proceed to a detailed calculation.

\section{Vacuum and the Gap Equation}
Write the ``classical" fields at $\mbit{J}=0$ as
\begin{eqnarray}
\mbit{\phi } = (m, \mbit{\Sigma}_\pi )  \ ; \qquad  \mbit{\Sigma}_\pi \equiv 
\left\{
\begin{array}{cc}
\pi \  ,	& D=4 \ ;	\\
(\pi_1, \pi_2)    \ ,  	& D=3 \ ;
 \end{array}
  \right.  \ , 
\end{eqnarray}
and 
\begin{eqnarray}
\tilde{\mbit{\phi }} \equiv (m + \varepsilon , \mbit{\Sigma}_\pi ) \ , 
\end{eqnarray}
to study the vacuum condition (\ref{VacCond}) 
on the tree part of the effective potential (\ref{effectivePotential}),
\begin{eqnarray}
{\cal V}_0 & = & \frac{1}{2 \lambda }\left( m^2 + \mbit{\Sigma}_\pi^2 \right)
  - \frac{1}{VT} {\rm Tr}\ln 
\left( \dslpar +  \tilde{\mbit{\phi }} \cdot \mbit{\Gamma}  
\right)  \nonumber \\
& = & \frac{1}{2 \lambda }\left( m^2 + \mbit{\Sigma}_\pi^2 \right) 
-  \frac{{\rm tr}{\bf I}}{2}\int \frac{d^D p}{(2 \pi )^D} 
\ln \left( p^2 + \tilde{\mbit{\phi }}^2  \right)  \ , 
\end{eqnarray}
giving 
\begin{eqnarray}
\frac{\partial {\cal V}_0}{\partial m} & = & 0 = \frac{m}{\lambda } 
  - 4 (m + \varepsilon) \int \frac{d^D p}{(2 \pi )^D}  \frac{1}{p^2 
+\tilde{\mbit{\phi }}^2} \  ,   \\ 
\frac{\partial {\cal V}_0}{\partial \mbit{\Sigma}_\pi} & = & 
0 = \frac{\mbit{\Sigma}_\pi}{\lambda } 
  - 4 \mbit{\Sigma}_\pi \int \frac{d^D p}{(2 \pi )^D}  
\frac{1}{p^2 +\tilde{\mbit{\phi }}^2} \  .  \label{TreeSigma}
\end{eqnarray}
(Recall ${\rm tr}{\bf I}= 4$ for both $D=4$ and $D=3$.) 
Therefore the solution is
\begin{eqnarray}
& & \mbit{\Sigma}_\pi=0 \ , \label{GapSigma} \\
& & \frac{m}{\lambda } 
  =  4 (m + \varepsilon) \int \frac{d^D p}{(2 \pi )^D}  
\frac{1}{p^2 +(m + \varepsilon)^2}  \ , \label{GapEQoneloop1}
\end{eqnarray}
whose second relation, called the gap equation in the tree order, reads
\begin{eqnarray}
\frac{1}{\lambda_D } =  \frac{\sqrt{x}}{\sqrt{x} - \epsilon} g_D^{(0)}(x)  \ ,
   \label{GapEQoneloop2}
\end{eqnarray}
where
\begin{eqnarray}
\epsilon \equiv \frac{\varepsilon }{\Lambda }  \ , \qquad   
x \equiv \frac{ (m + \varepsilon)^2  }{\Lambda^2}   \ , \quad  
\left( \epsilon^2  \leq  x \leq 1 \right)     \  , 
\end{eqnarray}
\begin{eqnarray}
g_4^{(0)}(x) & \! \! \equiv \! \! &   
1 - x \ln \left( \frac{1+x}{x} \right)     \ ,  \label{4Dg} \\ 
g_3^{(0)}(x) & \! \!  \equiv \! \! &   
2 \left[ 1 - \sqrt{x} \tan^{-1}  \frac{1}{\sqrt{x}}  \right]        \ ,  
\label{3Dg}
\end{eqnarray}
and
\begin{eqnarray}
\lambda_D  \equiv  
\frac{\lambda \Lambda^{D-2} }{4^{D/2-1} \pi^{D/2}{\mathrm \Gamma}(D/2)} \ ,  
\label{newlambda} 
\end{eqnarray} 
with ${\mathrm \Gamma}(D/2)$ being the gamma function; ${\mathrm \Gamma}(2) = 1 , {\mathrm \Gamma}(3/2) = \sqrt{\pi }/2$. In view of eq.(\ref{GapEQoneloop2}), 
\begin{eqnarray}
\lambda \stackrel{\sqrt{x} \mapsto \epsilon }{=} \mathcal{O}(\sqrt{x} - \epsilon)  \ ,  
\end{eqnarray}
which implies a trivial fact that mass is $\varepsilon$ even in the free theory $\lambda =0$.

The one-loop part ${\cal V}_1$ of the effective potential (\ref{effectivePotential}) reads
\begin{eqnarray}
{\cal V}_1  = \frac{1}{2N}  {\bf tr}  \int \frac{d^D p}{(2\pi)^D} 
\ln  \left( \frac{\bf I}{\lambda} + {\bf \Pi}(p) \right)   \ ,  
\label{OneLoopPart}
\end{eqnarray}
where {\bf tr} should be taken for the spinorial space as well as the $\mbit{\Sigma}$. In eq.(\ref{OneLoopPart}) the argument of the logarithm is nothing but a two-point function of the auxiliary fields, therefore, we call ${\bf \Pi}$ a vacuum polarization matrix,
\begin{eqnarray}
{\rm  \Pi}_{ab}(p) & \equiv & \int \frac{d^D l}{(2\pi)^D} {\rm tr}
\left( \frac{1}{i (\dsll+ \dslp /2) + \tilde{\mbit{\phi }} \cdot \mbit{\Gamma}  } {\Gamma }_a \frac{1}{i (\dsll- \dslp /2) + \tilde{\mbit{\phi }} \cdot \mbit{\Gamma} } {\Gamma}_b \right)  \nonumber  \\ 
& = &  4 \int \frac{d^D l}{(2\pi)^D} \frac{1}{\left[ (l + p/2)^2 + \tilde{\mbit{\phi }}^2 \right] \left[ (l - p/2)^2 + \tilde{\mbit{\phi }}^2 \right]}  \nonumber   \\ 
& \times & \hspace{-4mm}  
  \left\{
  \begin{array}{c}
 \left(
  \begin{array}{cc}
\displaystyle{ - (l^2 - \frac{p^2}{4}) + (m + \varepsilon )^2 - \pi^2 }	&   2 (m + \varepsilon )\pi 	\\
 \noalign{\vspace{1mm}}
  2 (m + \varepsilon )\pi   	&  \displaystyle{-(l^2 - \frac{p^2}{4}) - (m + \varepsilon )^2 + \pi^2 }
  \end{array}
  \right) \ ; \  D = 4   \ ; 	\\
\noalign{\vspace{8mm}}
\left(
 \begin{array}{cc}
 \displaystyle{ - (l^2 - \frac{p^2}{4}) + (m + \varepsilon )^2 
  - \Sigma_\pi^2 } 	&   2 (m + \varepsilon )\pi_1 	\\ 
 2(m + \varepsilon )\pi_1	&  \displaystyle{-(l^2 - \frac{p^2}{4}) - (m + \varepsilon )^2 + \pi_1^2 -  \pi_2^2}	\\ 
 \noalign{\vspace{1mm}}
 2 (m + \varepsilon )\pi_2  	&   2 \pi_1 \pi_2
 \end{array}  \right.         \\
  \noalign{\vspace{2mm}}  
\left. \begin{array}{c}  2(m + \varepsilon )\pi_2	\\
\noalign{\vspace{1mm}}
   2 \pi_1 \pi_2	\\
  \displaystyle{-(l^2 - \frac{p^2}{4}) - (m + \varepsilon )^2 - \pi_1^2 + \pi_2^2}    \end{array}  \right)        
 \ ; \    D= 3  \ .    \end{array}  
  \right.     \label{VacPolar0}
\end{eqnarray}
By noting
\begin{eqnarray}
{\rm tr} \left[ \left( \frac{\bf I}{\lambda } + {\bf \Pi} \right)^{-1} \cdot \frac{\partial {\bf \Pi}}{\partial ({\Sigma }_\pi)^a}  \right]  = ({\Sigma }_\pi)^a(\cdots ) \ ,
\end{eqnarray}
and the tree relation (\ref{TreeSigma}), the vacuum condition for $\mbit{\Sigma }_\pi$ up to the one-loop is
\begin{eqnarray}
\left.\frac{\partial {\cal V}}{\partial \mbit{\Sigma }_\pi}\right|_{\mbit{J}=0}  = \mbit{\Sigma }_\pi (\frac{1}{\lambda } + \cdots )   \ , 
\end{eqnarray}
with ${\cal V} = {\cal V}_0 + {\cal V}_1$, allowing us to choose $\mbit{\Sigma}_\pi =0$ as a vacuum. The second fact is that in the one-loop part of the gap equation,
\begin{eqnarray}
\left.\frac{\partial {\cal V}_1}{\partial m}\right|_{\mbit{J}=0,\mbit{\Sigma}_\pi=0} 
\end{eqnarray}
we can utilize the tree result (\ref{GapEQoneloop1}).

Therefore in the gap equation we put $\mbit{\Sigma}_\pi =0$ in the expression (\ref{VacPolar0}) to find the diagonal matrix
\begin{eqnarray}
\left. {\bf \Pi }\right|_{\mbit{\Sigma}_\pi=0} = \left\{
             \begin{array}{ll}
     \left(
     \begin{array}{cc}
  {\rm \Pi }_1    	& 0	\\
     0 	& {\rm \Pi }_2
     \end{array}
     \right)       \ ;   	&  D=4 	\\
     \noalign{\vspace{3mm}}
   \left(
   \begin{array}{ccc}
   {\rm \Pi }_1 	&   0  &  0	\\
    0	&   {\rm \Pi }_2   &  0 \\ 
    0	&   0  &   {\rm \Pi }_2
   \end{array}
   \right)         \ ;   	&    D=3 
             \end{array}
             \right.    \ , 
\end{eqnarray}
with
\begin{eqnarray}
 \left\{
       \begin{array}{c}
    {\rm  \Pi}_1   	\\
    {\rm  \Pi}_2   
       \end{array}
       \right\} 
& \equiv  & 4  \int \frac{d^D l}{(2 \pi )^D} \frac{1}{\left\{ (l+p/2)^2 + (m + \varepsilon)^2 \right\}\left\{ (l-p/2)^2 + (m + \varepsilon)^2 \right\}  }  \nonumber  \\ 
& & \hspace{20mm}\times   
 \left\{
       \begin{array}{l}
 \displaystyle{ - \left( l^2 - \frac{p^2}{4} \right)  + (m + \varepsilon)^2  }                      \\
\displaystyle{  - \left( l^2 - \frac{p^2}{4} \right)  - (m + \varepsilon)^2  }    
     \end{array}   \right\} 
\ .   \label{VacPolar}
\end{eqnarray}
Write 
\begin{eqnarray}
l^2 - \frac{p^2}{4} = \frac{\left( l + p/2 \right)^2 + \left( l - p/2 \right)^2 }{2}   -\frac{ p^2}{2}  \ ,  \nonumber 
\end{eqnarray}
then
\begin{eqnarray}
 \left\{
       \begin{array}{c}
    {\rm  \Pi}_1   	\\
    {\rm  \Pi}_2   
       \end{array}
       \right\}  & \! \! =  \! \! & -2 \left[ \int \frac{d^D l}{(2 \pi)^D} \frac{1}{\left( l + p/2 \right)^2 + (m + \varepsilon)^2} + \int \frac{d^D l}{(2 \pi)^D} \frac{1}{\left( l - p/2 \right)^2 + (m + \varepsilon)^2} \right]  \nonumber \\ 
&& +  8  \int \frac{d^D l}{(2 \pi)^D} \frac{  \left\{
\begin{array}{c}
 (m + \varepsilon)^2 + p^2/4  \\
p^2/4 
\end{array}
 \right\} }{\left\{ \left( l + p/2 \right)^2 + (m + \varepsilon)^2 \right\}\left\{ \left( l - p/2 \right)^2 + (m + \varepsilon)^2 \right\}  }  \ .
\label{VacuumPlolarization}
\end{eqnarray}
Shifting the momentum (although we are in a cutoff world), we obtain
\begin{eqnarray}
 \left\{
       \begin{array}{c}
    {\rm  \Pi}_1   	\\
    {\rm  \Pi}_2   
       \end{array}
       \right\}  & = & - 4 \int \frac{d^D l}{(2 \pi)^D} \frac{1}{ l^2 + (m + \varepsilon)^2}  \nonumber  \\
& & + 8 \int_{-1/2}^{1/2} dt \int \frac{d^D l}{(2 \pi)^D} \frac{ \left\{
\begin{array}{c}
\displaystyle{  (m + \varepsilon)^2 + p^2/4 }   \\
\displaystyle{  p^2/4  }                                                   \end{array} \right\}}{\left[ l^2 + p^2(1/4 -t^2) + (m + \varepsilon)^2 \right] ^2  }  
  \ . \label{VacuumPolarizationShifted}
\end{eqnarray}
Using the tree result of the gap equation (\ref{GapEQoneloop1}), we obtain
\begin{eqnarray}
&&\frac{1}{\lambda } +  \left\{
       \begin{array}{c}
    {\rm  \Pi}_1   	\\
    {\rm  \Pi}_2   
       \end{array}
       \right\}  \nonumber \\
 &=& \frac{1}{\lambda}- 4 \int \frac{d^D l}{(2 \pi)^D} \frac{1}{ l^2 
+ (m + \varepsilon)^2}
   + 8 \int_{-1/2}^{1/2} dt \int \frac{d^D l}{(2 \pi)^D} \frac{ \left\{
       \begin{array}{c}
         \displaystyle{  (m+\varepsilon)^2 + p^2/4 }   \\
         \displaystyle{  p^2/4  }
       \end{array}
     \right\}}{\left[ l^2 + p^2(1/4 -t^2) + (m+\varepsilon)^2 \right] ^2  }
   \nonumber \\
   &=& \frac{1}{\lambda }  \left( 1 - \lambda_D g_D^{(0)}(x) + 4 \lambda_D
   q_D(x,s)
     \left\{ \! \! \!
\begin{array}{c}
  x + s    \\
s
  \end{array}
   \right\} \right)   \ , \label{2LoopLog}
\end{eqnarray}
where
\begin{eqnarray}
q_4(x,s) & \! \!  \equiv \! \!   & \frac{1}{2}\left[ \ln \frac{1 + 
x}{x}  + \frac{1+2x+2s}{2\sqrt{s(1 + x + s)}} \ln \frac{\sqrt{1 + x + 
s} + \sqrt{s}}{\sqrt{1 + x + s} - \sqrt{s}}  \right.   \nonumber \\
& & \hspace{50mm} \left. - \sqrt{\frac{x + s}{s}}\ln \frac{\sqrt{x + 
s} + \sqrt{s}}{\sqrt{x + s} - \sqrt{s}}  \right]  \ ; \\
q_3(x,s) & \! \!  \equiv \! \!   &  \int_{0}^{1} dt \frac{1}{\sqrt{x 
+ s(1 -t^2)}}    \tan^{-1}\frac{1}{\sqrt{x + s(1 -t^2)}}  \nonumber \\
& & \hspace{30mm}   - \frac{1}{2\sqrt{s(1+x+s)}} \ln \left( 
\frac{\sqrt{1 + x + s} + \sqrt{s}}{\sqrt{1 + x + s} - \sqrt{s}} 
\right)   \ ;
\end{eqnarray}
with 
\begin{eqnarray}
s \equiv \frac{p^2}{4 \Lambda^2}  \ ; \qquad  0 \leq s \leq  \frac{1}{4} \ ,  
\end{eqnarray}
and $\lambda_D$ being given by eq.(\ref{newlambda}).

It should be noticed that in eq.(\ref{OneLoopPart}) the $\mbit{\Gamma}_5$ part of $1/\lambda + {\rm \Pi}_2$ , the (inverse) propagator of pions, vanishes when $s \mapsto 0$ as well as $\epsilon \mapsto 0$; that is, pions are the massless Nambu-Goldstone particle(s). The quantity $\epsilon$, therefore, plays a role of an infrared cutoff.

\begin{figure}[h]
$$
\epsfxsize=8cm \epsfbox{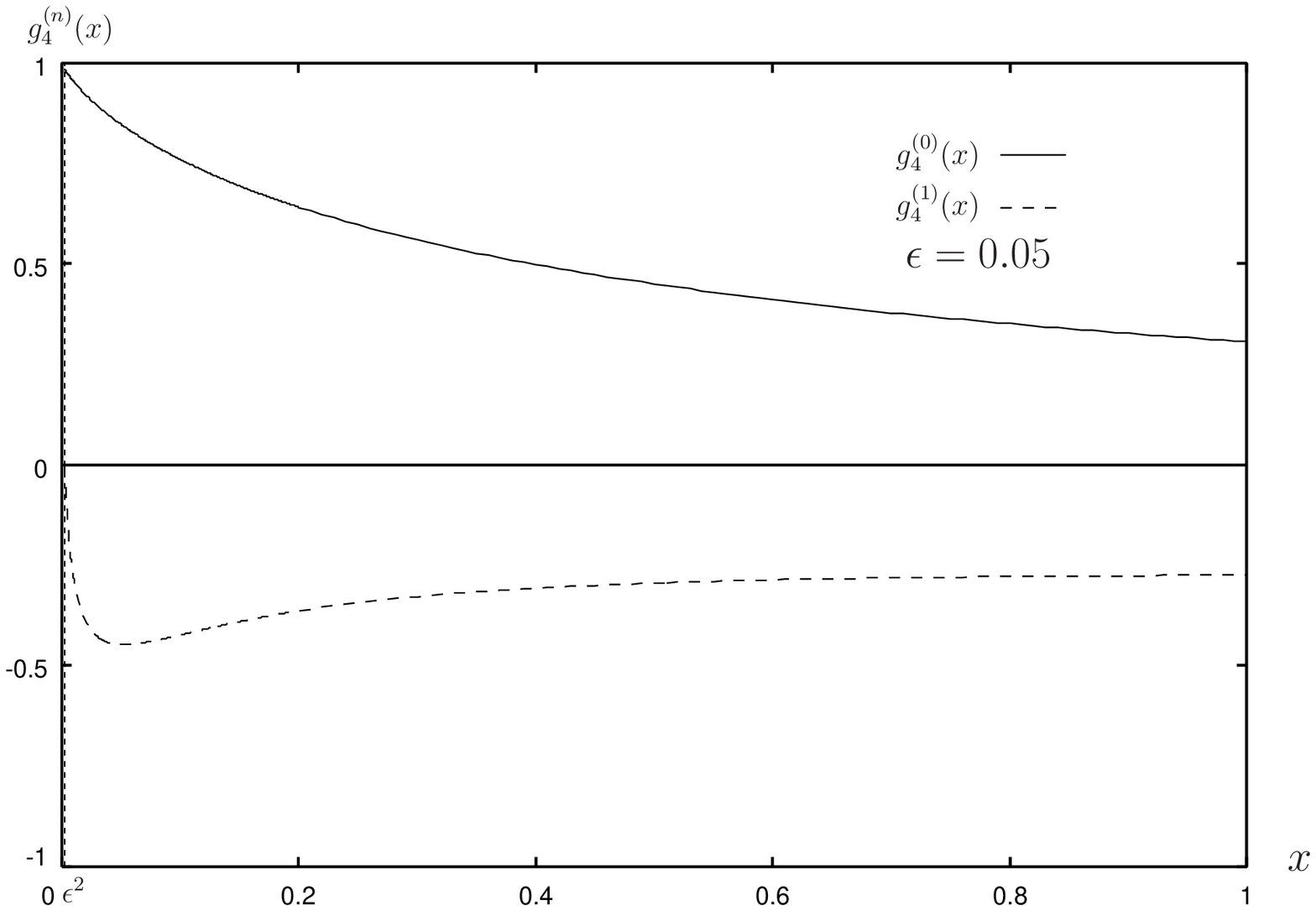} \  \epsfxsize=8cm \epsfbox{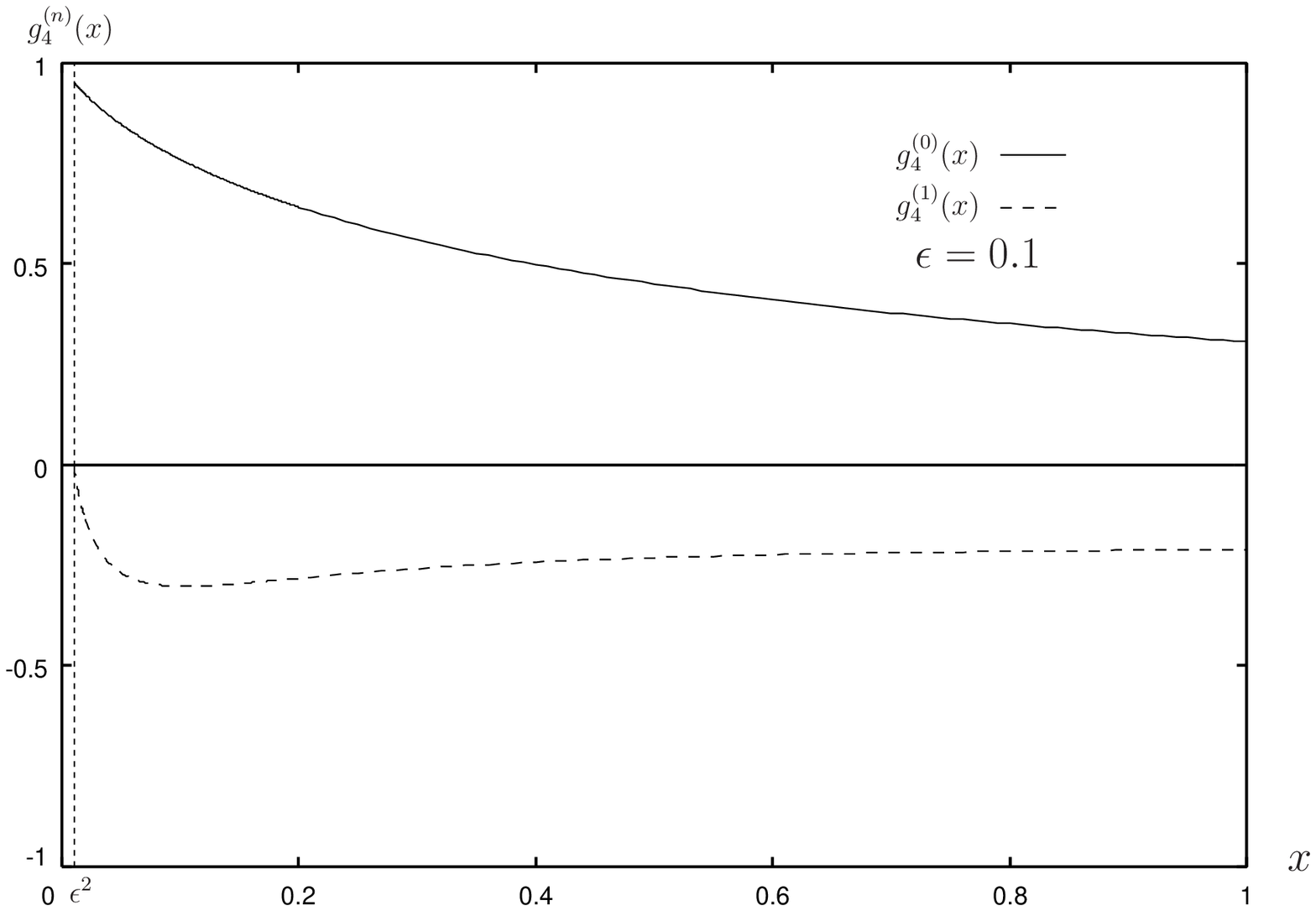}  
$$

\vspace{-5mm}

\caption{$g^{(0)}_4(x)$(a solid line) and $g^{(1)}_4(x)$(a dotted line) are plotted with $\epsilon$ being set $0.05$(left) and $0.1$(right). $x \equiv (m + \varepsilon )^2/ \Lambda^2$ and $\epsilon \equiv \varepsilon / \Lambda $. Note that $g^{(0)}_4(x)$ is positive but $g^{(1)}_4(x)$ is negative everywhere.}
\label{fig:gvalue4}
\end{figure}

\begin{figure}[h]
$$
\epsfxsize=8cm \epsfbox{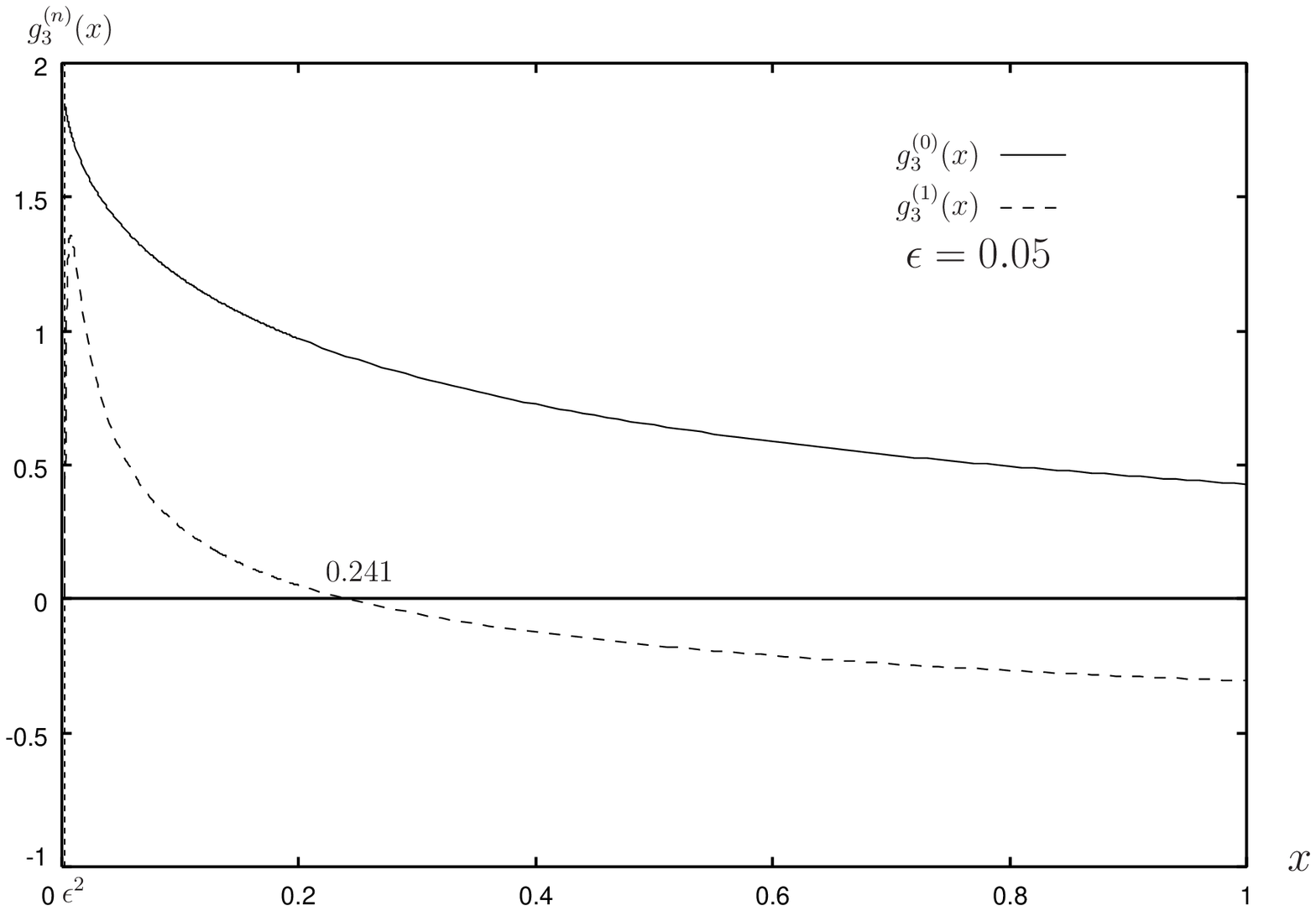} \  \epsfxsize=8cm \epsfbox{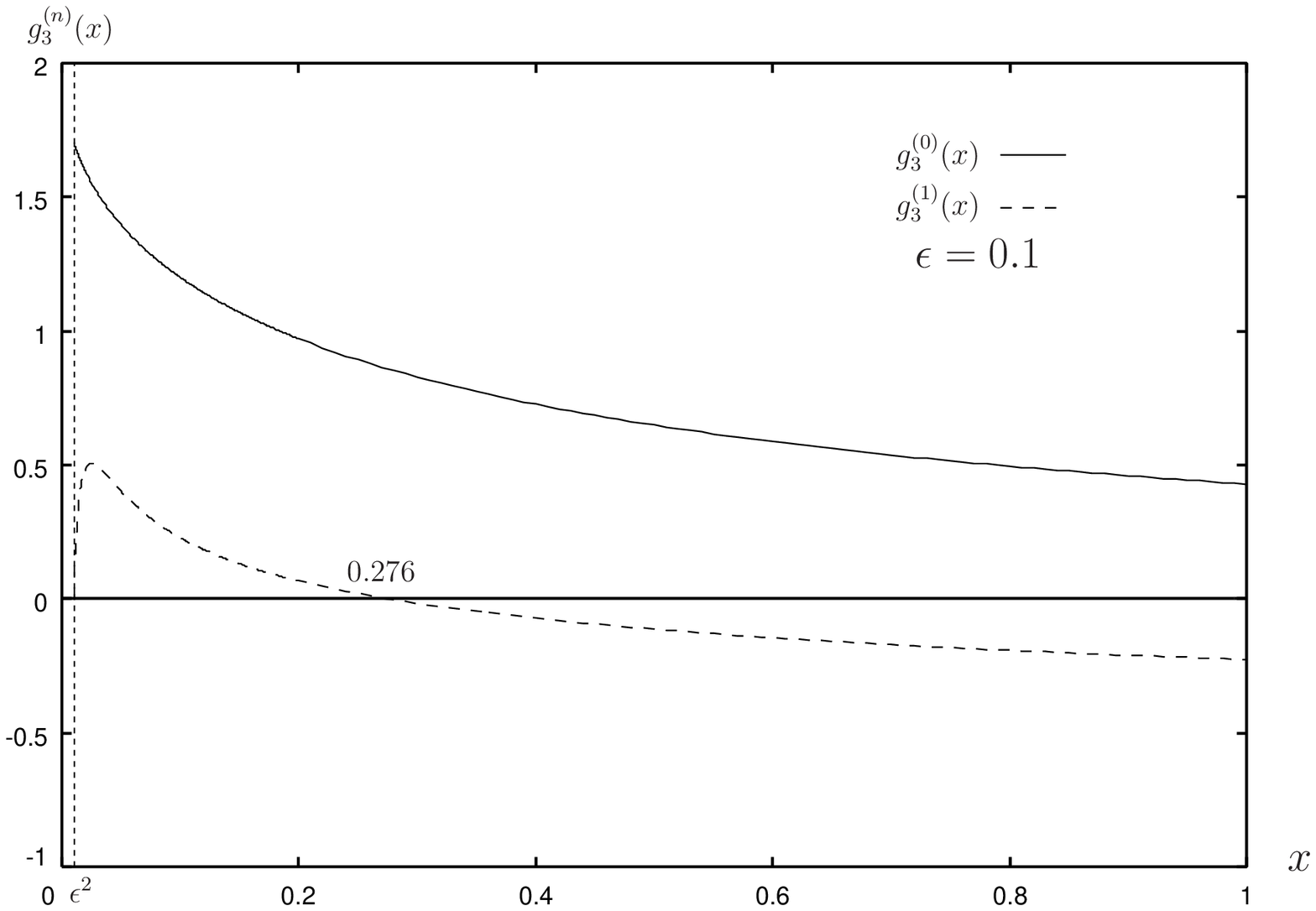}
$$

\vspace{-5mm}

\caption{$g^{(0)}_3(x)$(a solid line) and $g^{(1)}_3(x)$(a dotted line) are plotted with $\epsilon$ being set $0.05$(left) and $0.1$(right). Note that $g^{(1)}_3(x)$ has a zero at $x=0.241(\epsilon=0.05)$ or at $x=0.276(\epsilon=0.1)$.}
\label{fig:gvalue3}
\end{figure}

The one-loop part of the gap equation is derived from a restricted (one-loop) effective potential obtained  by putting $\mbit{\Sigma}_\pi=0$ in eq.(\ref{OneLoopPart}),
\begin{eqnarray}
 \left.{\cal V }_1 \right|_{\mbit{\Sigma}_\pi = 0} = \frac{1}{2 N} \frac{\Lambda^D}{ \pi^{D/2}{\mathrm \Gamma}(D/2)}   \int_{0}^{1/4} ds s^{(D-2)/2} Q_D(x,s) +   \mbox{ $x$-independent terms}  \ , 
 \end{eqnarray}
where
\begin{eqnarray}
  Q_4(x,s) & \equiv & \ln \left( 1 - \lambda_4 g_4^{(0)}(x) + 4 \lambda_4 
    (x+s)q_4(x,s) \right) \nonumber \\ 
  &&  +   \ln \left( 1 - \lambda_4 g_4^{(0)}(x) + 4 
    \lambda_4 s q_4(x,s) \right)  \ ,       \\
  \noalign{\vspace{1mm}}
  Q_3(x,s) & \equiv & \ln \left( 1 - \lambda_3 g_3^{(0)}(x) + 4 \lambda_3 
    (x+s)q_3(x,s) \right)  \nonumber \\
  &&  +   2\ln \left( 1 - \lambda_3 g_3^{(0)}(x) + 4 
    \lambda_3 s q_3(x,s) \right)
      \ ,
\end{eqnarray}
to give
\begin{eqnarray}
\left. \frac{\partial {\cal V}_1}{\partial x} \right|_{\mbit{\Sigma}_\pi = 0, \mbit{J}=0}& \! \! =\! \! & \frac{1}{2N}
\frac{\Lambda^D}{ \pi^{D/2}{\mathrm \Gamma}(D/2)} \int_{0}^{1/4} ds s^{(D-2)/2}  \frac{\partial Q_D(x,s)}{ \partial x} \nonumber \\ 
 & \equiv  & - \frac{1}{2 N} \frac{ \Lambda^D}{ 4^{D/2-1} \pi^{D/2}{\mathrm \Gamma}(D/2)}g_D^{(1)}(x)   \ ; 
\end{eqnarray}
where
\begin{eqnarray}
& & \hspace{-13mm} g_4^{(1)}(x)   \equiv  - 4 \int_{0}^{1/4}  \! \!  \! \! ds s 
  \left( 
    \frac{-\lambda_4 g_{4,x}^{(0)} 
      + 4 \lambda_4 ( q_4 +(x+s) q_{4,x} ) }{1-\lambda_4 g_4^{(0)} 
      + 4 \lambda_4  (x+s) q_4  } 
    + \frac{-\lambda_4 g_{4,x}^{(0)}
      + 4 \lambda_4 s q_{4,x}  }{1-\lambda_4 g_4^{(0)} 
      + 4 \lambda_4 s q_4  }  
  \right)  \nonumber \\
  && \hspace{-10mm} =  - 4 (\sqrt{x}-\epsilon) \int_{0}^{1/4}  \! \!  \! \! ds s \left( 
    \frac{- g_{4,x}^{(0)} 
      + 4 ( q_4 +(x+s) q_{4,x} ) }{\epsilon g_4^{(0)} 
      + 4 (\sqrt{x}-\epsilon) (x+s) q_4  } 
    + \frac{- g_{4,x}^{(0)}
      + 4 s q_{4,x}  }{\epsilon g_4^{(0)} 
      + 4 (\sqrt{x}-\epsilon) s q_4  }  
  \right),
\label{1loop4dim}    \\
 & & \hspace{-13mm} g_3^{(1)}(x) \equiv - 2 \int_{0}^{1/4}  \! \!   \! \! ds \sqrt{s} \left( 
    \frac{-\lambda_3 g_{3,x}^{(0)} 
      + 4 \lambda_3 ( q_3 +(x+s) q_{3,x} ) }{1-\lambda_3 g_3^{(0)} 
      + 4 \lambda_3  (x+s) q_3  } 
    + 2 \frac{-\lambda_3 g_{3,x}^{(0)}
      + 4 \lambda_3 s q_{3,x}  }{1-\lambda_3 g_3^{(0)} 
      + 4 \lambda_3 s q_3  }  
  \right)  \nonumber \\
 & & \hspace{-10mm} = - 2 (\sqrt{x}-\epsilon) \int_{0}^{1/4}  \! \! \! \! ds \sqrt{s} \left( 
    \frac{- g_{3,x}^{(0)} 
      + 4 ( q_3 +(x+s) q_{3,x} ) }{\epsilon g_3^{(0)} 
      + 4 (\sqrt{x}-\epsilon) (x+s) q_3  } 
    + 2 \frac{- g_{3,x}^{(0)}
      + 4 s q_{3,x}  }{\epsilon g_3^{(0)} 
      + 4 (\sqrt{x}-\epsilon) s q_3  }  
  \right) \ , 
\label{1loop3dim}
\end{eqnarray}
with 
\begin{eqnarray}
q_{D, x} \equiv \frac{\partial q_D}{ \partial x }  \ . 
\end{eqnarray}
In eqs.(\ref{1loop4dim}) and (\ref{1loop3dim}), 
we have replaced $\lambda_D$ by the tree value, 
$(\sqrt{x} - \epsilon )/ (\sqrt{x}  g_D^{(0)}(x))$ in the first expressions. 
In Figures \ref{fig:gvalue4} and \ref{fig:gvalue3} 
we show the shape of $g^{(0)}_D(x)$ as well as $g^{(1)}_D(x)$ 
for $ 0 \leq x \equiv (m + \varepsilon )^2/ \Lambda^2  \leq 1$. 
We choose two cases for the infrared cutoff 
$\epsilon \equiv \varepsilon / \Lambda $  as $0.05$ and $0.1$. 

The gap equation, up to the one-loop order of the auxiliary fields, therefore, is found
\begin{eqnarray}
\frac{1}{ \lambda_D  } = \frac{\sqrt{x}}{\sqrt{x} - \epsilon } \left( g_D^{(0)}(x) + \frac{1}{N}g_D^{(1)}(x) \right)    \ , \label{FullGapEq}
\end{eqnarray}
whose right hand side diverges at $\sqrt{x}=\epsilon$, implying that $\lambda_D=0$. As was discussed before, this is physically reasonable since we have a current quark mass $\varepsilon$ even in a free $\lambda =0$ theory. In view of eq.(\ref{FullGapEq}), an important role of the current quark mass $\varepsilon$ under the loop expansion is recognized: when $\varepsilon = 0 \ (\epsilon = 0)$ the right hand side reads
\begin{eqnarray*}
 \left. \left( g_D^{(0)}(x) + \frac{1}{N}g_D^{(1)}(x) \right) \right|_{\epsilon = 0} \ , 
\end{eqnarray*}
whose second term becomes infinite. (Note that in eqs.(\ref{1loop4dim}) and (\ref{1loop3dim}), $q_{D, x}$ is singular at $x=0$ under $\epsilon=0$.) Thus the second term surpasses the first, causing a breakdown of the loop expansion. On the contrary, if $\epsilon \neq 0$, the second term is much smaller in the dangerous region $\sqrt{x} \sim \epsilon $ owing to the factor $\sqrt{x} - \epsilon $ in front of the integrals in eqs.(\ref{1loop4dim}) and (\ref{1loop3dim}). 
\begin{figure}[h]
$$
\epsfxsize=8cm \epsfbox{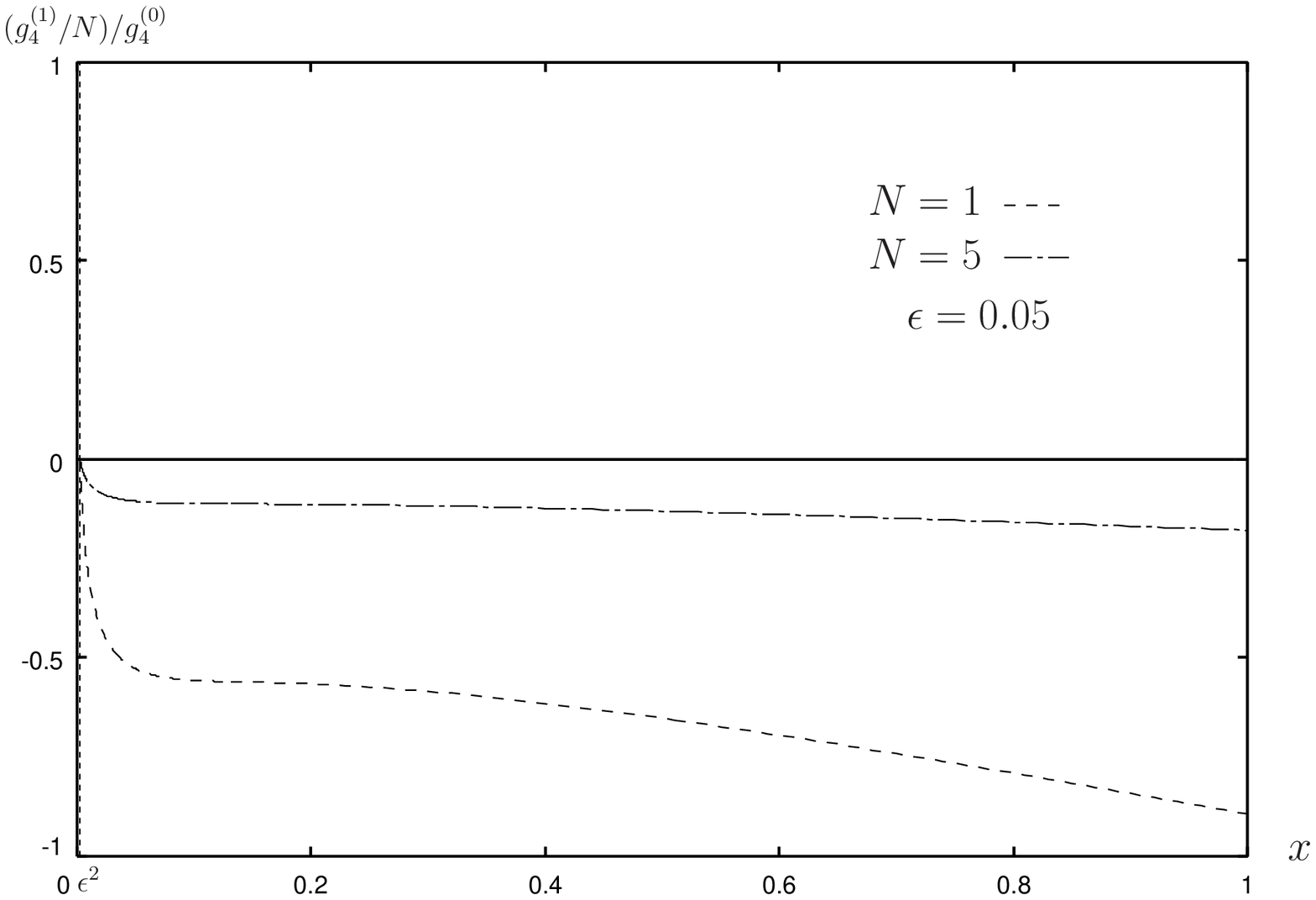} \  \epsfxsize=8cm \epsfbox{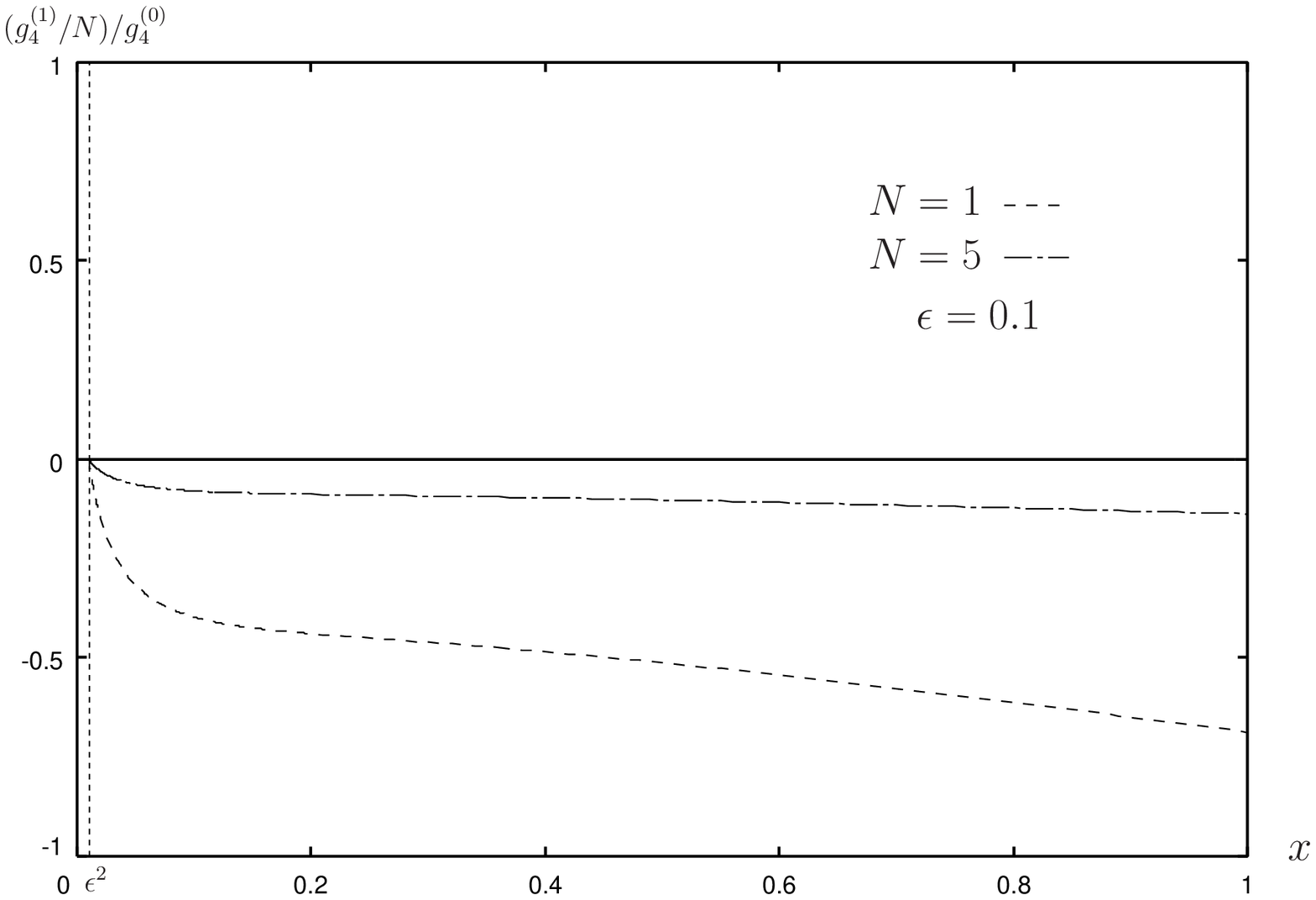}  
$$

\vspace{-5mm}

\caption{The ratio $(g_4^{(1)}(x)/N)/g_4^{(0)}(x)$: the left graphs are for $\epsilon = 0.05$ and the right for $0.1$. The dotted line stands for $N=1$ and the dash-dotted line for $N=5$. It is recognized that the ratio remains less than unity even in $N=1$.}
\label{fig:Ratio4}
\end{figure}

\begin{figure}[h]
$$
\epsfxsize=8cm \epsfbox{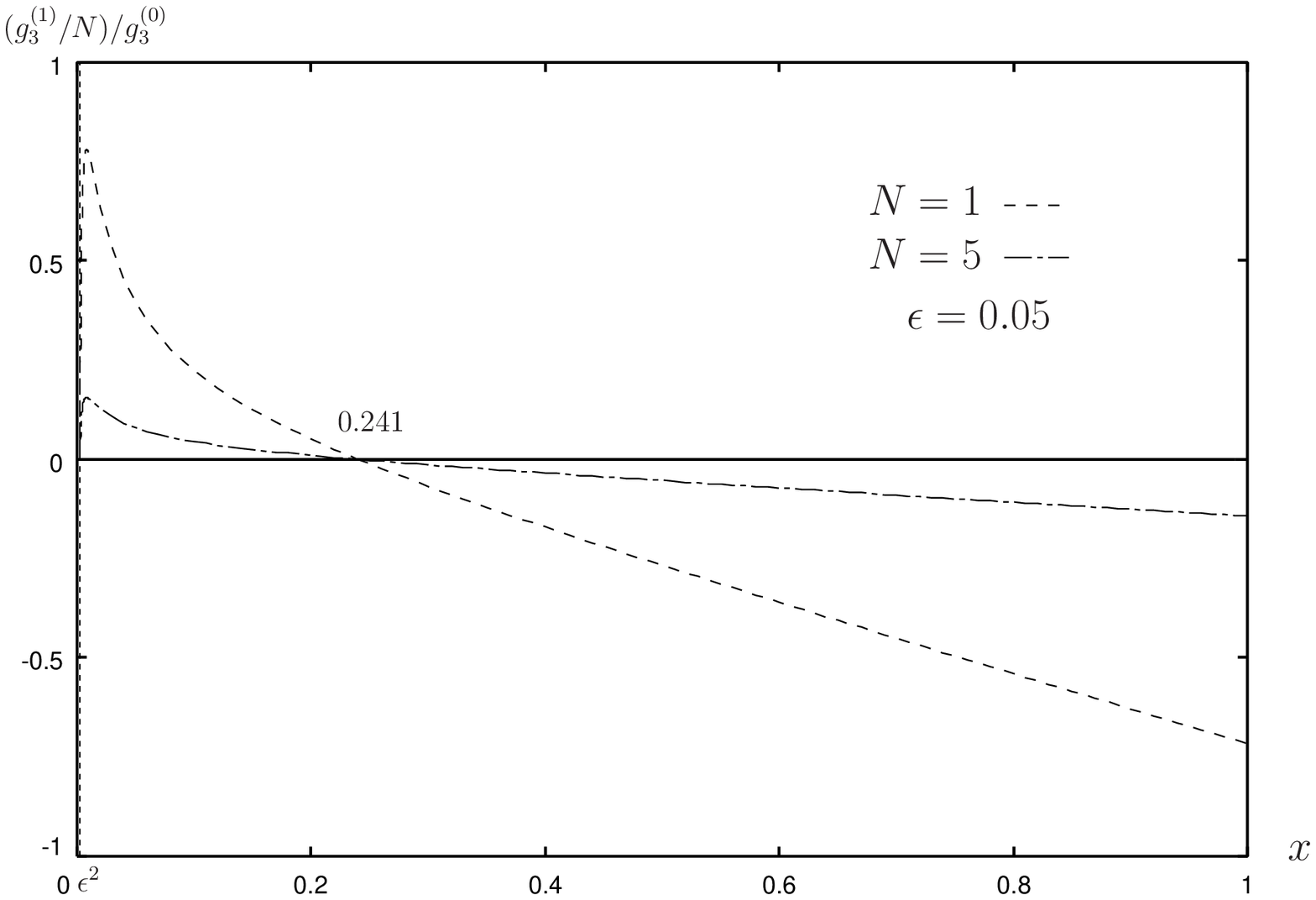} \   \epsfxsize=8cm \epsfbox{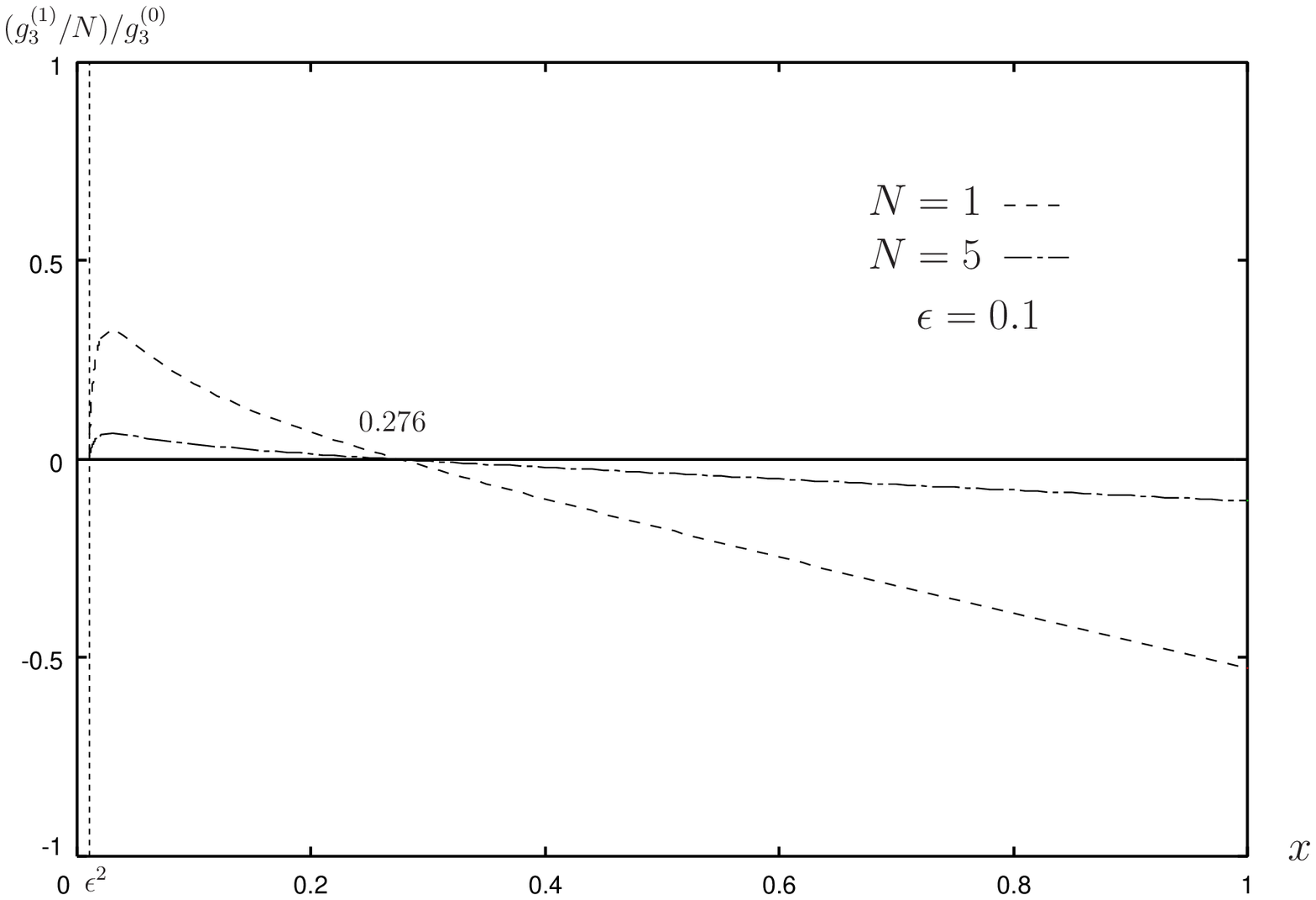}
$$

\vspace{-5mm}

\caption{The ratio $(g_3^{(1)}(x)/N)/g_3^{(0)}(x)$: the left graphs are for $\epsilon = 0.05$ and the right for $0.1$. The dotted line stands for $N=1$ and the dash-dotted line for $N=5$. It is recognized that the ratio remains less than unity even in $N=1$.}
\label{fig:Ratio3}
\end{figure}
In Figures \ref{fig:Ratio4} and \ref{fig:Ratio3}, 
we plot the ratios $(g_D^{(1)}/N)/g_D^{(0)}$ in $D=4$ and $D=3$, 
respectively. It is recognized that even in $N=1$(the dotted line) 
the loop expansion is legitimated; since the ratio remains less than unity. 
It is also shown that the smaller $\epsilon$ becomes the greater the ratio goes. 
The critical values that cause the ratio to exceed unity 
under $N=1$ are then found such that $\epsilon=0.0326$ at $x=1$ in $D=4$ and $\epsilon=0.040$ at $x=0.0052$ in $D=3$. 
Those of $\epsilon$ can be put smaller when $N$ goes larger.

\begin{figure}[h]
$$
\epsfxsize=8cm \epsfbox{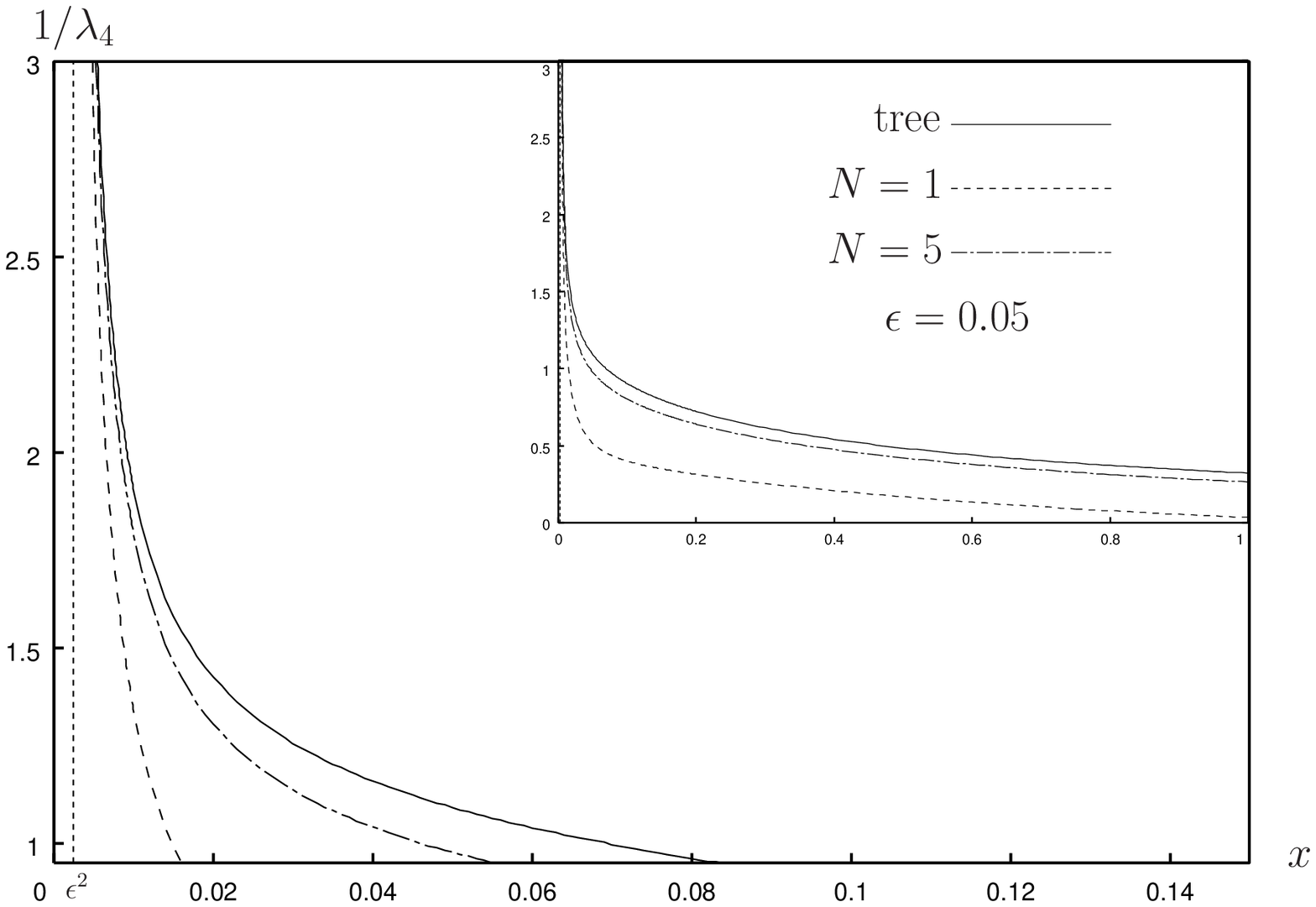}  \  \epsfxsize=8cm \epsfbox{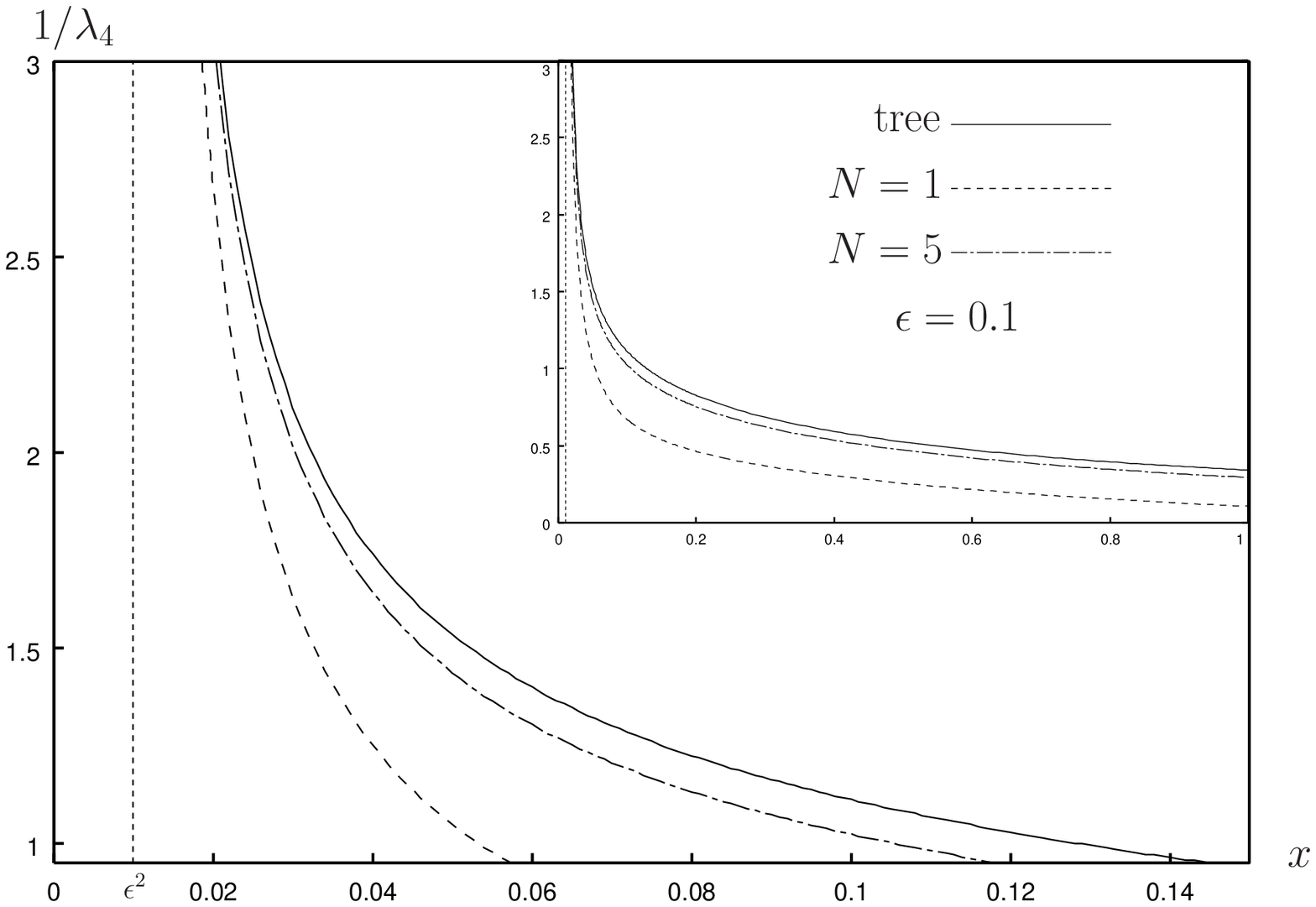} 
$$

\vspace{-5mm}

\caption{The right hand side of the gap equation in $D=4$: the solid line designates the tree order or $N= \infty$ , while the dotted and the dash-dotted ones include the one-loop effects with $N=1$ and $N=5$, respectively. $\epsilon$ is put 0.05(left) and 0.1(right).  Graphs with $0 \leq x < 0.15$ are shown inside which the whole shapes $0 \leq x \leq 1$ are given. $x$ is recognized as a monotone increasing function of $N$ with respect to a (supposed) horizontal line, namely a fixed four-Fermi coupling.}
\label{fig:D=4}
\end{figure}

\begin{figure}[h]

$$
\epsfxsize=8cm \epsfbox{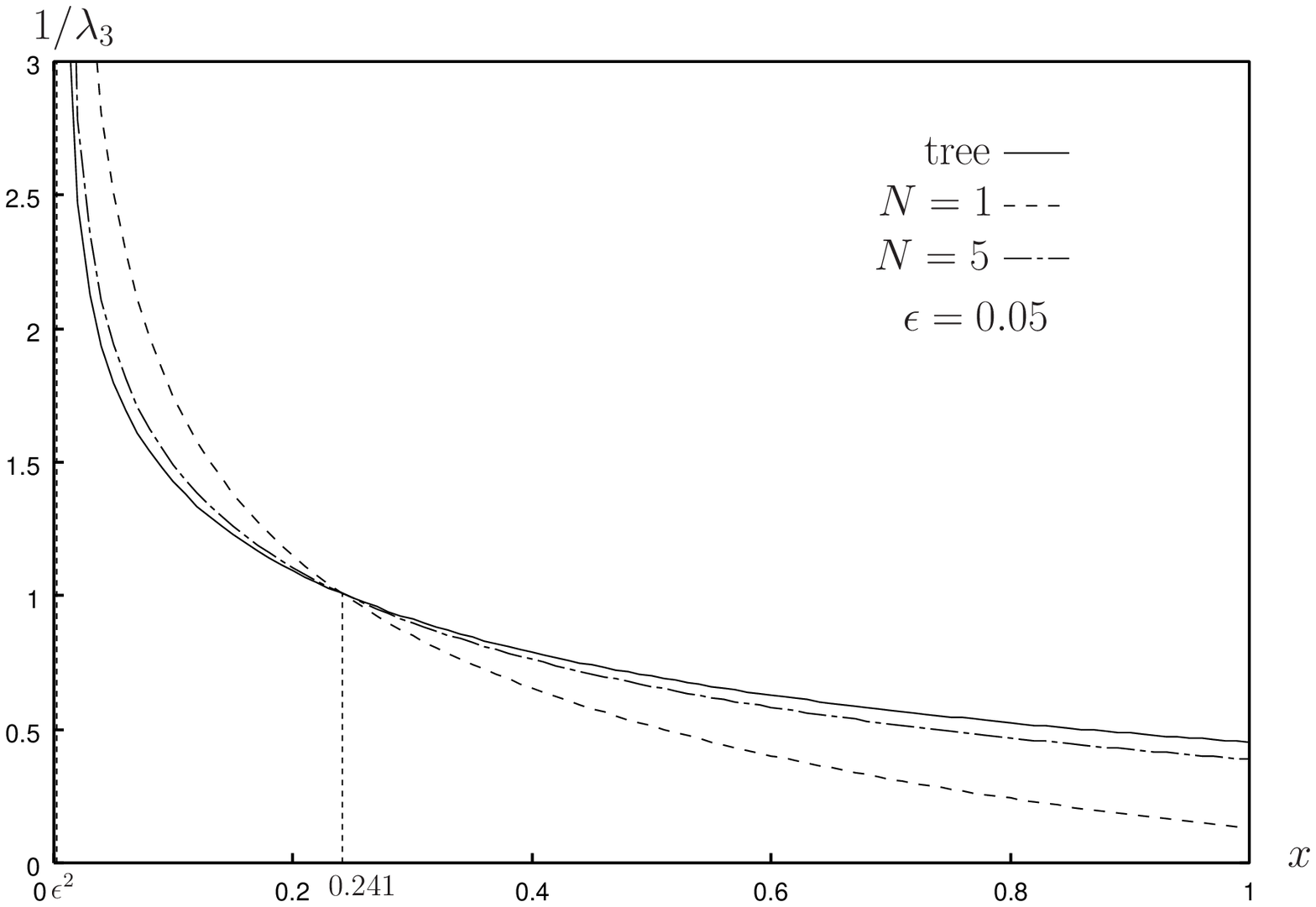} \   \epsfxsize=8cm \epsfbox{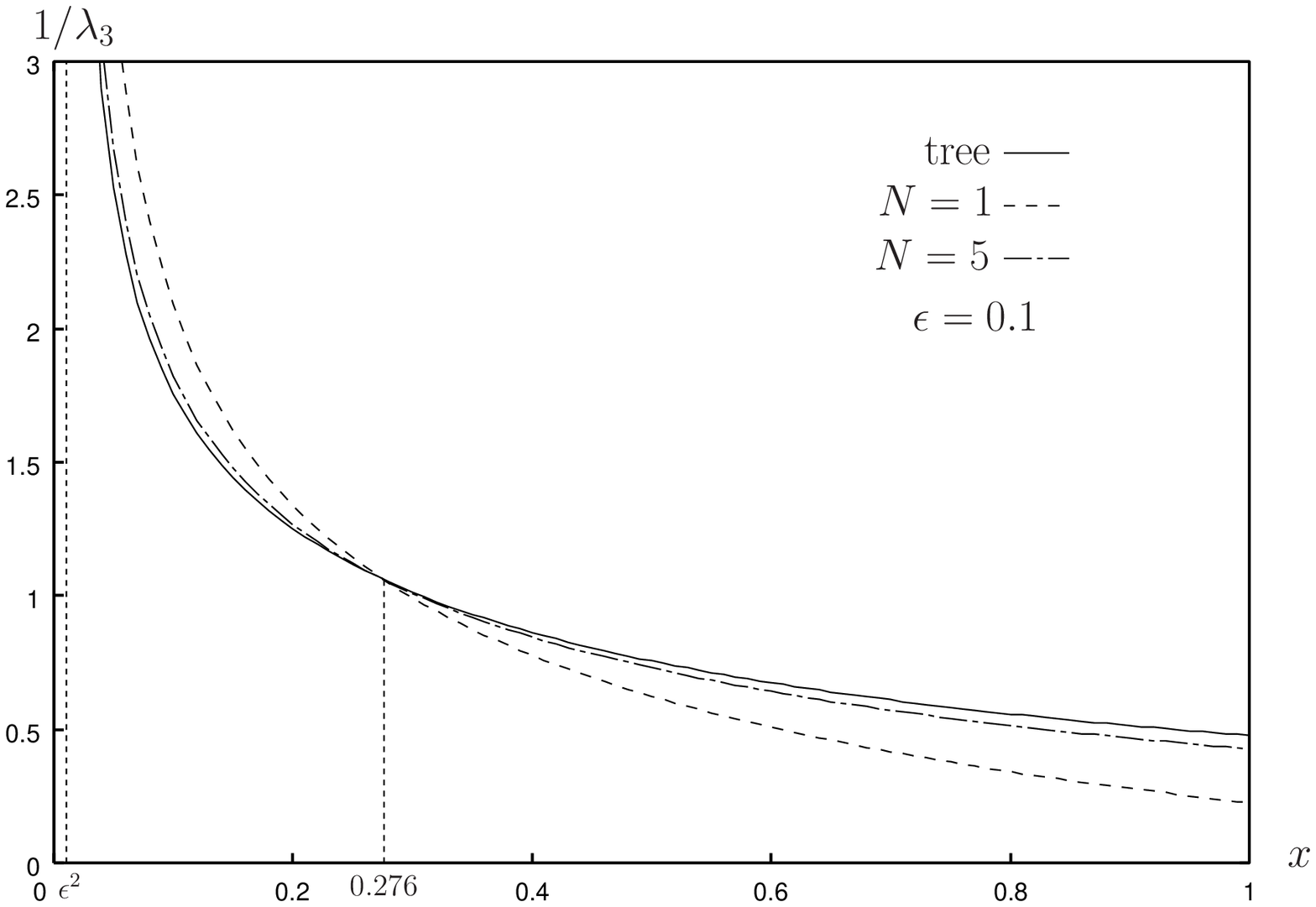}
$$

\vspace{-5mm}

\caption{The right hand side of the gap equation in $D=3$: the solid line designates the tree order or $N= \infty$ , while the dotted and the dash-dotted ones include the one-loop effects with $N=1$ and $N=5$ respectively. $\epsilon$ is put 0.05(left) and 0.1(right). In $x \leq  0.241(0.276)$ $x$ is a monotone decreasing function of $N$ for a fixed coupling, but on the contrary in $x >  0.241(0.276)$ that is an  increasing function.}
\label{fig:D=3}
\end{figure}

Finally we plot the right hand side of the gap equation (\ref{FullGapEq}) 
for $D=4$ and $D=3$ in the Figures \ref{fig:D=4} and \ref{fig:D=3}. 
The horizontal line again stands for the value of 
$x = (m + \varepsilon )^2/ \Lambda^2$ and the vertical line 
for $1/\lambda_D$. The solid line designates that of 
$\sqrt{x} g_D^{(0)}(x)/ (\sqrt{x} - \epsilon )$, namely, 
the tree or $N=\infty$ case. The dotted and the dash-dotted lines 
include $\mathcal{O}(1/N)$ contribution with $N=1$ and $N=5$, respectively. 
We have set $\epsilon = 0.05 $(left graphs) and $\epsilon = 0.1 $
(right graphs). In $D=4$ magnified figures for $0 \leq x < 0.15$ 
are shown together with the whole plots inside those.

It is seen that for a fixed four-Fermi coupling $\lambda_D$, that is, with respect to a (supposed) horizontal line, 
mass $x$ is a monotone increasing function of $N$ in $D=4$, 
but in $D=3$ the dependence is not so simple because of zeros 
in $g^{(1)}_3(x)$ (see the Figures \ref{fig:gvalue4} 
and \ref{fig:gvalue3}); $x$ is monotone decreasing(increasing) in a small(large) mass or a four-Fermi coupling region. Physically speaking, due to quantum effects, 
$\chi$SB is restored in $D=4$ at any coupling. Meanwhile in $D=3$ it is restored(enhanced) in a strong(weak) coupling region.



\section{Discussion}

In this paper we have examined the higher order(= quantum) effect 
of auxiliary fields to the gap equation in the NJL model. 
Contrary to the observation by Kleinert and Bossche\cite{rf:kleinert}, 
we find that auxiliary fields still play a significant role 
under the nonvanishing current quark mass $\varepsilon$. 
``Pions can still survive" in the NJL model. 
We cannot put the intrinsic fermion mass into zero 
but at the order of $\Lambda/100$ when $N=1$, 
in order to overcome the infrared divergences 
and to ensure the loop expansion. 
In $D=4$ the dynamical mass $x$ is a monotone increasing function of $N$ 
at any fixed four-Fermi coupling constant. 
But in $D=3$ $x$ is monotone decreasing(increasing) 
under a small(large) mass or coupling region. 
In other words, dynamical mass shrinks by means of quantum effects 
in the strong coupling regime of $D=3$ as well as in $D=4$. 
On the contrary it swells under the weak coupling regime in $D=3$. 
We have already encountered a similar situation in $D=3$ in the
reference \cite{rf:IKT2}; dynamical mass is a complicated function 
of the magnitude of the background magnetic fields(MBMF) under the 
influence of quantum gluons. In a small mass or coupling region, 
it is a monotone increasing function of MBMF, 
but decreasing function in a larger mass or coupling region.

In this way, we recognize that the auxiliary field method for the NJL model 
can survive with an infrared cutoff. 
The power of the auxiliary field method is shown 
in ref.\cite{rf:kashiwa}, using $0$- and $1$-dimensional examples. 
Cases have been, however, only for bose ones so that analysis 
for fermionic models is necessary. 
The $0$-dimensional fermionic model, 
the Grassmann integration model, 
is studied to fulfill our expectation that inclusion 
of higher-loop effects of auxiliary fields 
improves a result much better\cite{rf:KS}.
The $1$-dimensional, quantum mechanical case is now under study.

\vspace{10mm}

\noindent{\Large \bf Acknowledgements}

\vspace{5mm}

\noindent 
This work is supported in part by Grant-in-Aid for Science Research 
from the Japan Ministry of Education, 
Science and Culture; 12640280 and 13135217.

\end{document}